\documentclass{pasj00}
\draft

\author{%
Mikio \textsc{Morii},\altaffilmark{1}
Mutsumi \textsc{Sugizaki},\altaffilmark{2}
Nobuyuki \textsc{Kawai},\altaffilmark{1}
Motoko \textsc{Serino},\altaffilmark{2}
Takayuki \textsc{Yamamoto},\altaffilmark{2, 3}\\
Ryuichi \textsc{Usui},\altaffilmark{1}
Arata \textsc{Daikyuji},\altaffilmark{4}
Ken \textsc{Ebisawa},\altaffilmark{5}
Satoshi \textsc{Eguchi},\altaffilmark{6}
Kazuo \textsc{Hiroi},\altaffilmark{6}\\
Masaki \textsc{Ishikawa},\altaffilmark{7}
Naoki \textsc{Isobe},\altaffilmark{6}
Kazuyoshi \textsc{Kawasaki},\altaffilmark{8}
Masashi \textsc{Kimura},\altaffilmark{9}
Hiroki \textsc{Kitayama},\altaffilmark{9}\\
Mitsuhiro \textsc{Kohama},\altaffilmark{8}
Takanori \textsc{Matsumura},\altaffilmark{10} 
Masaru \textsc{Matsuoka},\altaffilmark{2}
Tatehiro \textsc{Mihara},\altaffilmark{2}\\
Yujin E. \textsc{Nakagawa},\altaffilmark{11}
Satoshi \textsc{Nakahira},\altaffilmark{12}
Motoki \textsc{Nakajima},\altaffilmark{13}
Hitoshi \textsc{Negoro},\altaffilmark{3}\\
Hiroshi \textsc{Ozawa},\altaffilmark{3}
Megumi \textsc{Shidatsu},\altaffilmark{6}
Tetsuya \textsc{Sootome},\altaffilmark{2}
Kousuke \textsc{Sugimori},\altaffilmark{1}
Fumitoshi \textsc{Suwa},\altaffilmark{3}\\
Hiroshi \textsc{Tomida},\altaffilmark{8}
Yohko \textsc{Tsuboi},\altaffilmark{10}
Hiroshi \textsc{Tsunemi},\altaffilmark{9}
Yoshihiro \textsc{Ueda},\altaffilmark{6}
Shiro \textsc{Ueno},\altaffilmark{8}\\
Akiko \textsc{Uzawa},\altaffilmark{10}
Kazutaka \textsc{Yamaoka},\altaffilmark{12}
Kyohei \textsc{Yamazaki},\altaffilmark{10}
and
Atsumasa \textsc{Yoshida}\altaffilmark{12}
}

\affil{
$^1$Department of Physics, Tokyo Institute of Technology, Ookayama 2-12-1, Meguro-ku,\\
Tokyo 152-8551, Japan\\
morii@hp.phys.titech.ac.jp\\
$^2$Coordinated Space Observation and Experiment Research Group, Institute of Physical\\
and Chemical Research (RIKEN), 2-1 Hirosawa, Wako, Saitama 351-0198, Japan\\
$^3$Department of Physics, Nihon University, 1-8-14 Surugadai, Chiyoda, Tokyo 101-8308, Japan\\
$^4$Department of Applied Physics, University of Miyazaki, 1-1 Gakuen Kibanadai-nishi,\\
Miyazaki, Miyazaki 889-2192, Japan\\
$^5$Department of Space Science Information Analysis, Institute of Space and Astronautical Science,\\
Japan Aerospace Exploration Agency, 3-1-1 Yoshino-dai, Chuo-ku, Sagamihara, Kanagawa 252-5210, Japan\\
$^6$Department of Astronomy, Kyoto University, Oiwake-cho, Sakyo-ku, Kyoto 606-8502, Japan\\
$^7$School of Physical Science, Space and Astronautical Science, The graduate University \\
for Advanced Studies (Sokendai), Yoshinodai 3-1-1, Chuo-ku, Sagamihara, Kanagawa 252-5210, Japan\\
$^8$ISS Science Project Office, Institute of Space and Astronautical Science,\\
Japan Aerospace Exploration Agency, 2-1-1 Sengen, Tsukuba, Ibaraki 305-8505, Japan\\
$^9$Department of Earth and Space Science, Osaka University, 1-1 Machikaneyama, Toyonaka,\\
Osaka 560-0043, Japan\\
$^{10}$Department of Physics, Faculty of Science and Engineering, Chuo University, 1-13-27 Kasuga,\\
Bunkyo-ku, Tokyo 112-8551, Japan\\
$^{11}$High Energy Astrophysics Laboratory, Institute of Physical and Chemical Research (RIKEN),\\
2-1 Hirosawa, Wako, Saitama 351-0198, Japan\\
$^{12}$Department of Physics and Mathematics, Aoyama Gakuin University, 5-10-1 Fuchinobe, Chuo-ku,\\
Sagamihara, Kanagawa 252-5258, Japan\\
$^{13}$School of Dentistry at Matsudo, Nihon University, 2-870-1 Sakaecho-nishi, Matsudo,\\
Chiba 271-8587, Japan
}

\title{MAXI GSC monitoring of the Crab nebula and pulsar during
the GeV gamma-ray flare in September 2010}

\Received{2011/04/19}
\Accepted{2011/07/20}


\SetRunningHead{Morii et al.}{Crab}

\begin{document}

\maketitle

\begin{abstract}
 
We report on the MAXI GSC X-ray monitoring of the Crab nebula and pulsar
during the GeV gamma-ray flare for the period of 2010 September 18$-$24 (MJD 55457$-$55463)
detected by AGILE and Fermi-LAT.
There were no significant variations on the pulse phase averaged 
and pulsed fluxes during the gamma-ray flare
on time scales from 0.5 to 5 days.
The pulse profile also showed no significant change during this period.
The upper limits on the variations of the pulse phase averaged 
and pulsed fluxes for the period MJD 55457.5$-$55462.5 in the 4$-$10 keV band are
derived to be 1 and 19\%, respectively, at the 90\% confidence limit of the statistical uncertainty.
The lack of variations in the pulsed component over the multi-wavelength range
(radio, X-ray, hard X-ray, and gamma-ray) supports 
not the pulsar but the nebular origin for the gamma-ray flare.
\end{abstract}

\section{Introduction}

The Crab nebula has been the standard candle in high energy X-ray and gamma-ray astronomy.
The flux and spectrum in these energy ranges have been expected to be steady over years.
Surprisingly, AGILE and Fermi-LAT reported a flare for the period of 2010 September 18$-$24
in the GeV gamma-ray energy range \citep{Tavani+2010, Buehler+2010, Tavani+2011, Abdo+2011}.
The first half of the flare (September 18$-$21) was detected by both AGILE and Fermi-LAT,
while that of the second half (September 21$-$24) was detected only by Fermi-LAT.
The half-day binned light curve of Fermi-LAT exhibited
three sub-flares during these periods \citep{Balbo+2011}.

INTEGRAL observed the Crab nebula 
from 10:32 on September 12 to 12:48 on September 19 (UT) for calibration purposes,
which covered the first fifth of the gamma-ray flare.
It detected no significant flux increase in the 20$-$400 keV range \citep{Ferrigno+2010a}.
Swift BAT detected no variations
over the uncertainty of 5.5\% at the 1-$\sigma$ limit in the 15$-$50 keV range
for the period of 2010 September 19$-$21 \citep{Markwardt+2010}.
Radio observations of the Crab pulsar showed no evidence 
of a pulsar glitch and also no change
on the pulsed flux as well as the pulse profile \citep{Espinoza+2010}.
ARGO-YBJ reported an excess of events (4 $\sigma$) from a direction consistent with the Crab nebula,
corresponding to a flux about 3$-$4 times higher than usual,
at the median energy of about 1 TeV \citep{Aielli+2010}.
On the other hand, MAGIC and VERITAS reported no significant enhancement
in the TeV gamma-ray flux during the GeV flare \citep{Mariotti+2010, Ong+2010}.
The Swift XRT follow-up observation of 1-ks exposure starting on September 22 at 16:42 (UT),
which corresponds to the third sub-flare 
recognized in the Fermi-LAT light curve \citep{Balbo+2011},
showed no changes of flux, spectrum or pulse profile \citep{Evangelista+2010}.
Swift XRT detected no active AGN near the Crab nebula \citep{Heinke 2010}.
The INTEGRAL follow-up observation from 21:05 on September 22 to 11:12 on September 23 (UT),
approximately corresponding to the end of the flare, found no significant change
in the pulse profile in the 20$-$40 keV band \citep{Balbo+2011}.

After the gamma-ray flare, the RXTE PCA follow-up observation on September 24 
showed no changes of flux, spectrum nor pulse profile \citep{Shaposhnikov+2010}.
The follow-up observation in a near-infrared wavelength
on September 24 showed no variation
in the Crab pulsar in the J and H bands \citep{Kanbach+2010}.
The follow-up observations by Chandra on September 28 and the HST on October 2 revealed
an anvil feature close to the base of the pulsar
jet and elongated striation at the distant place
\citep{Tennant+2010, Ferrigno+2010b, Horns+2010, Tavani+2011}.
These features are thought to indicate a particle acceleration and the origin
of the GeV gamma-ray flare \citep{Tavani+2011}.
Two further GeV gamma-ray flare episodes,
detected by AGILE on 2007 October and by Fermi-LAT on 2009 February,
were reported \citep{Tavani+2011, Abdo+2011}.
None of these three gamma-ray flares showed any changes of the pulsed component
in the GeV energy range.

Another surprising thing is the long-term variability
of the Crab nebula in the hard X-ray bands reported by \citet{Wilson-Hodge+2011}.
The total nebula flux was found to decrease 
from 2008 August to 2010 July
and the fractional decline was larger in higher energy ranges.
On the other hand, the pulsed flux in the $3.2-35$ keV band of RXTE PCA
decreased steadily at
$\sim 0.2$\% yr$^{-1}$, consistent with the pulsar spin-down,
indicating that the observed X-ray variability
would originate not from the pulsar but from the nebula.

MAXI has been monitoring the Crab nebula
since the beginning of the mission on 2009 August 15 (UT),
which covered the whole gamma-ray flare in 2010 September. 
Here, we report on the MAXI monitoring of the pulse phase averaged and pulsed
fluxes of the Crab nebula, as well as the pulse profile.

\section{Observation}\label{sec: observation}

MAXI (Monitor of All-sky X-ray Image) is an X-ray all-sky monitor,
mounted on the Japanese Experiment Module - Exposed Facility 
of the International Space Station \citep{Matsuoka+2009}.
It carries two types of X-ray cameras:
the Gas Slit Camera (GSC; \cite{Mihara+2011, Sugizaki+2011}) for the 2$-$30 keV band
and Solid-state Slit Camera (SSC; \cite{Tsunemi+2010, Tomida+2011}) for the 0.5$-$12 keV band,
using gas proportional counters and X-ray CCDs, respectively.
Since the time resolution of the SSC ($5.6$ s)
is too long to detect the pulsation of the Crab pulsar,
we concentrate on the analysis of the GSC data in this paper.

The GSC scans almost all the sky every 92 minutes with a field of view (FoV) of
\timeform{1.5D} (FWHM) by \timeform{160D}.
The effective area for any source is calculated
according to the collimator transmission function of
a triangular shape during each scan transit
\citep{Sugizaki+2011, Morii+2006, Morii+2010}.
The spatial resolution of the GSC is approximately \timeform{1.5D} (FWHM).
The time resolution of the GSC is 50 $\mu$s.
We confirmed that the relative event time was stable
within the standard deviation of 0.2 ms
throughout the whole observation period.
The absolute time assignment was also confirmed as
accurate within the stability of the relative time,
by comparing the main peak phase of the Crab pulsar obtained 
by the GSC with that of the RXTE PCA observation \citep{Rots+2004}.


\section{Analysis and Results}\label{sec: analysis}

\subsection{Pulse phase averaged flux of the Crab nebula}\label{subsec: total flux}

We examined the variation of the entire Crab nebula flux averaged over the pulse phase.
We employed the same procedure to derive the Crab flux to
that in the effective-area calibration described in \citet{Sugizaki+2011}.
The source event data were extracted from the region of \timeform{1.6D} radius from
the Crab nebula.
The background level was estimated from the
data for the adjacent source-free sky region within the annulus with the
inner and outer radii of \timeform{1.6D} and \timeform{3.2D}, respectively,
where the region contaminated by the nearby bright source, namely A 0535+262, was excluded.
We used the data of the four anodes (C0, C3, C4, and C5) among the total of six (C0$-$C5)
for all the counters in this analysis 
because the other two were not calibrated well
for the complex energy-PHA responses \citep{Sugizaki+2011}.
Figure \ref{fig: lc} (top panel) shows the light curve obtained in the 4$-$10
keV band in 2.5-day time bins. Here, the binning boundaries were carefully chosen
to divide the flare period (MJD 55457.5$-$55462.5) into two periods ``A'' and ``B''
at MJD 55460 (see figure 2 in \cite{Balbo+2011}).
The flux for the entire flare period ``A+B'' and those for Periods A and B
are shown in table \ref{flux_values}.

The small fluctuations in the light curve are thought to be
not intrinsic to the Crab nebula, but due to the systematic uncertainties
in the GSC effective area calibration.
To estimate the systematic uncertainty, 
we fitted the light curve with linear functions,
the decline rate of which was either set free or
fixed to that determined with the RXTE PCA light curve 
from MJD 54690 to 55435 in the 2$-$15 keV band \citep{Wilson-Hodge+2011}.
The obtained parameters of the model functions are shown in table \ref{model_values},
where they are denoted as ``Average: linear (free)'' and ``Average: linear (fixed)'', respectively.
Assuming that the systematic uncertainty affects all the data bins uniformly
and is proportional to the flux,
we estimate the systematic uncertainty
by calculating the modified reduced chi-squared 
$\chi^2_{\rm red} = \frac{1}{\rm DOF}
\sum_i [(F_i - F_{\rm model}(t_i))^2] / [\sigma_{F_i}^2 + (r_{\rm syst} F_i)^2]$
for a tentative systematic uncertainty ($r_{\rm syst}$).
Here, $F_i$ and $\sigma_{F_i}$ are the flux and statistical error 
at the $i$-th time bin ($t_i$), respectively.
$F_{\rm model}$ and ${\rm DOF}$ are the model function for the flux
and the degree of freedom, respectively.
We then estimated the systematic uncertainty of 1-$\sigma$ level
by searching for the $r_{\rm syst}$ to make the reduced chi-squared unity.
The obtained systematic uncertainties were 2\% in both cases.
On the other hand, the maximum deviation from the best-fit functions
was 8\% among 118 time bins in both cases.
Since the latter deviation is statistically large in comparison
with the former uncertainty,
there are some peculiar bins which are subject to larger systematic uncertainty.

The flux variations observed during Periods A, B, and A+B are statistically  
consistent with the best fit linear model.
We calculated the upper limits on the variation of the flux 
during the flare, expressed by the excess ratio (\%)
to the best-fit model functions (table \ref{model_values}).
The statistical 90\% confidence level upper limits during Period A+B are 1.0 and 0.8\%
for the two linear model functions.
The values during Periods A and B are also shown in table \ref{model_values}.

To investigate the variability corresponding
to the sub-flares observed by Fermi-LAT \citep{Balbo+2011},
we made the 0.5-day binned light curve
in the 4$-$10 keV band during the flare interval
as shown in figure \ref{fig: lc_flare} (top panel).
The deviation from the best-fit linear function ``Average: linear (free)'' of table \ref{model_values}
is not statistically significant with a reduced chi-squared of 1.13 for 12 degree of freedom (DOF).

\subsection{Pulsed flux of the Crab pulsar deduced from sub-scan-duration data analysis}
\label{subsec: pulsed flux}

To measure the pulsed flux of the Crab pulsar, we analyzed the data by the following steps.
Since the background rate within the FoV
depends on the position in the detector,
we extracted events based on the detector coordinate.
We selected events within 5 mm from the position coincident with
the Crab nebula along the anode wires, which corresponds to about \timeform{2D}
on the sky.
We removed events from the scan period when
the instantaneous effective area of a GSC camera
was smaller than 1 cm$^2$.
This is because the systematic uncertainty and 
the signal-to-background ratio worsen under this condition.
The photon arrival times were corrected to the solar system barycenter by \texttt{mxbarycen},
the validity of which was confirmed by the timing calibration \citep{Sugizaki+2011}.
We chose the energy band of $4 - 10$ keV.
The events with corrected times were folded in the pulse period of the Crab pulsar
of the Jodrell Bank radio observatory \citep{Lyne+1993}.
We applied corrections of the effective area and exposure in this step.

The pulse profile during Period A+B is shown in
figure \ref{fig: pulse_flare}, where the phase zero 
corresponds to that of the first main pulse in radio \citep{Lyne+1993}.
To compare it with the normal pulse profile of the Crab pulsar, we made 
a template pulse profile $T(\phi_i)$ ($i = 1$st, $\cdots$, $N$-th phase bin; $N = 64$
and $\phi$ is the pulse phase.) by averaging the profile from 2009 December 13 to  2010 January 11,
in which the un-pulsed component was subtracted.
We fit the pulse profile during the gamma-ray flare to a model
$a \times T(\phi) + b$, where $a$ and $b$ represent
the scale factor of the pulsed component relative to the template
and the constant offset representing the background and the nebula component.
The best-fit model is shown as the solid line in figure \ref{fig: pulse_flare},
where the reduced chi-squared of the fit is 1.34 for 62 DOF.
We also performed the same analysis for the pulse profiles during
Periods A and B.
The reduced chi-squared of the fits are 0.92 and 1.35 for 62 DOF in these periods.
All the pulse profiles during Periods A, B and A+B
are consistent with the template pulse profile within the 99\% confidence limit.
%
%
From the pulse profile fitting, we also obtained the pulsed flux by
$a \sum_i T(\phi_i) / N$ 
 \footnote{Please note that the pulse profiles are not normalized to unity but
have the unit of counts cm$^{-2}$ s$^{-1}$. Therefore, this value becomes the pulsed flux.},
the results of which are shown in table \ref{flux_values}.
This method is free from background variation because
the background variation only affects the offset parameter $b$.

\begin{figure}
  \begin{center}
    \FigureFile(150mm,150mm){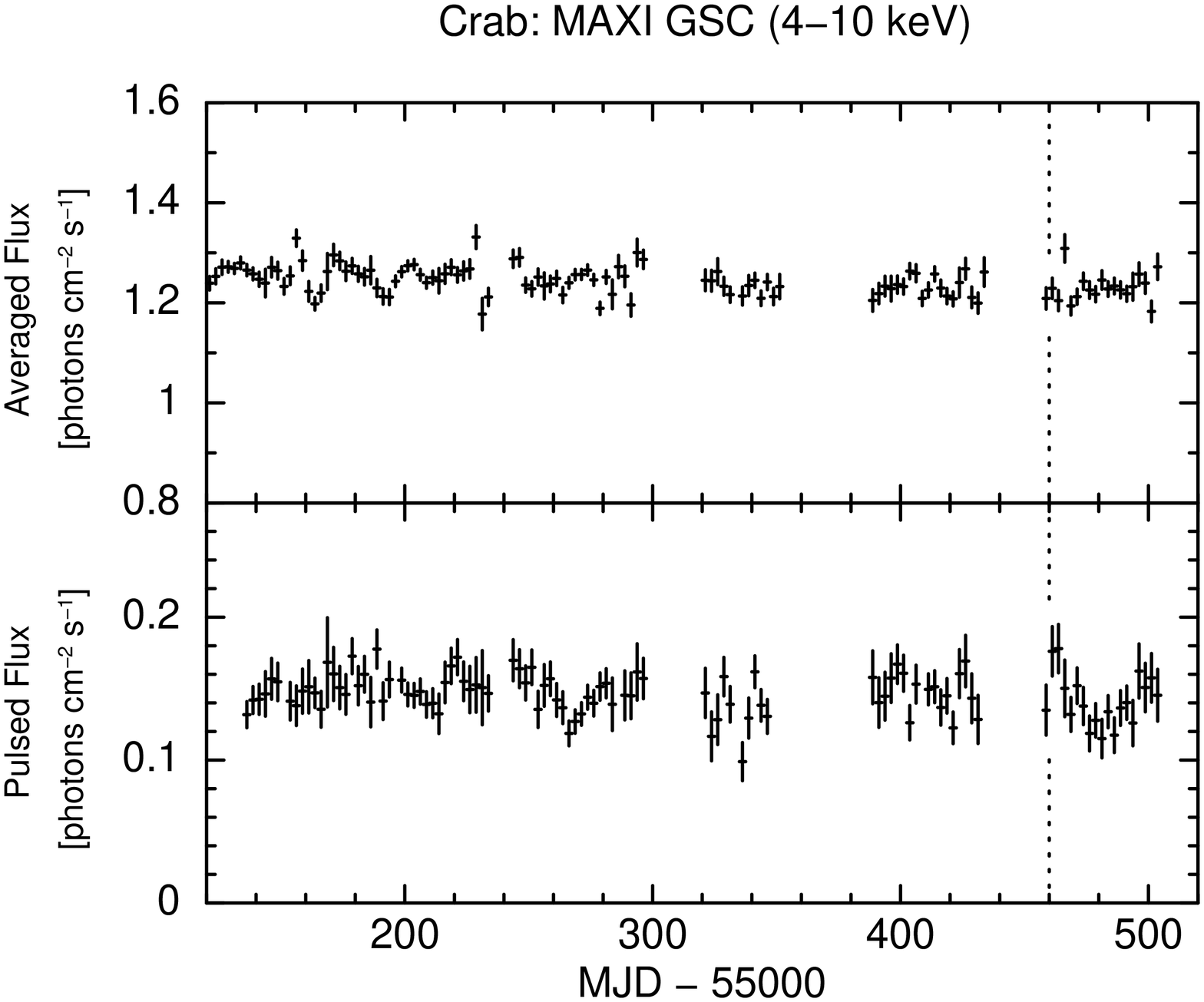}
  \end{center}
  \caption{GSC light curves of the pulse phase averaged flux
(top panel) and pulsed flux (bottom panel) of the Crab nebula
in the 4$-$10 keV band in 2.5-day time bins
from 2009 October 16 
to   2010 November 20
(MJD 55120$-$55520).
The horizontal and vertical axes are
shown in units of MJD minus 55000 (2009 June 18)
and photons cm$^{-2}$ s$^{-1}$, respectively.
The vertical error bars correspond to 1-$\sigma$ statistical errors.
The time center (2010 September 21; MJD 55460.0) of the period of the GeV gamma-ray flare
(MJD 55457.5$-$55462.5) is denoted by a vertical dotted line.
}\label{fig: lc}
\end{figure}

\begin{figure}
  \begin{center}
    \FigureFile(150mm,150mm){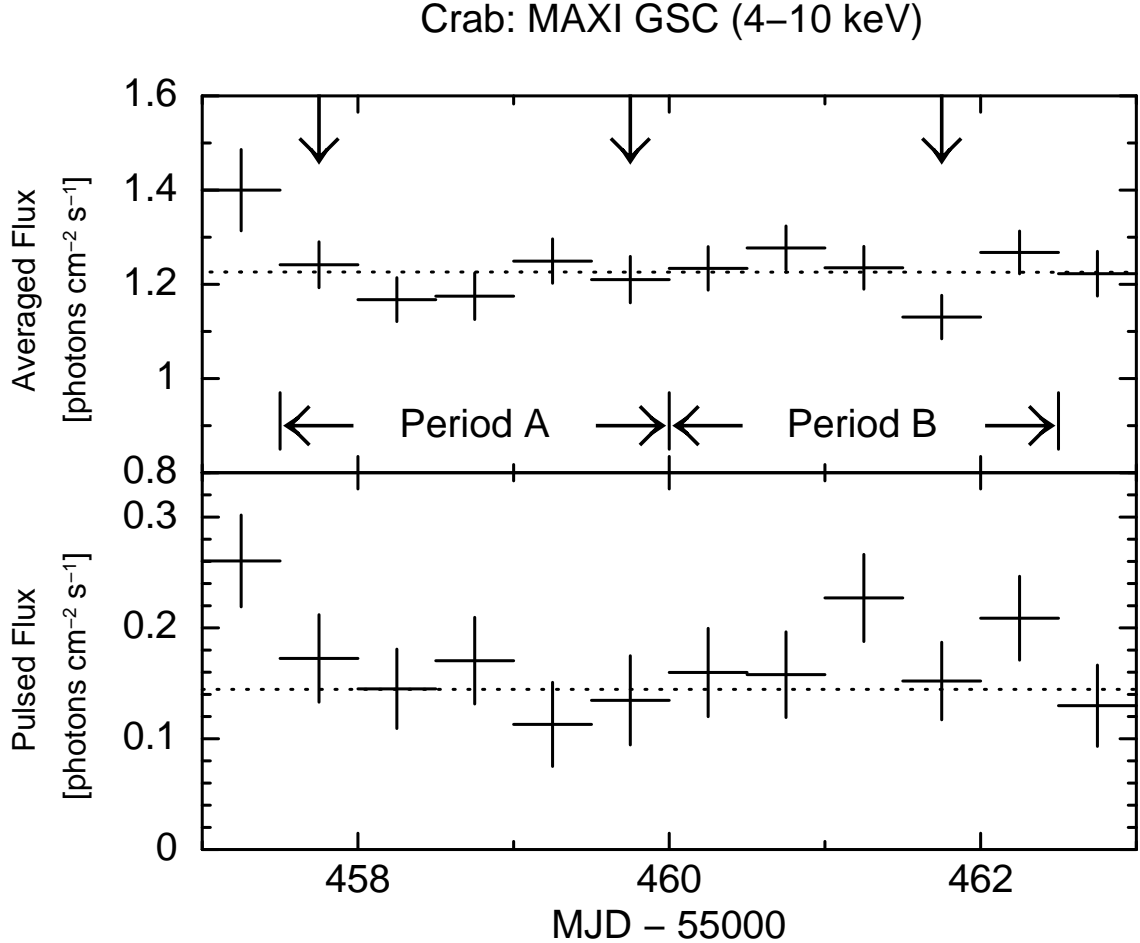}
  \end{center}
  \caption{Detailed light curve of figure \ref{fig: lc}
around the GeV gamma-ray flare from 2010 September 18 to 24
(MJD 55457$-$55463) in 0.5-day time bin.
The horizontal and vertical axes are shown in the same units as figure \ref{fig: lc}.
The vertical error bars correspond to 1-$\sigma$ statistical errors.
The peak times of the gamma-ray sub-flares \citep{Balbo+2011} are designated
in three downward arrows.
In the top and bottom panels, the dotted lines show
the linear function ``Average: linear (free)'' and
the constant value ``Pulsed: const'' with the best-fit parameters
of table \ref{model_values}, respectively.
  }\label{fig: lc_flare}
\end{figure}

\begin{figure}
  \begin{center}
    \FigureFile(150mm,150mm){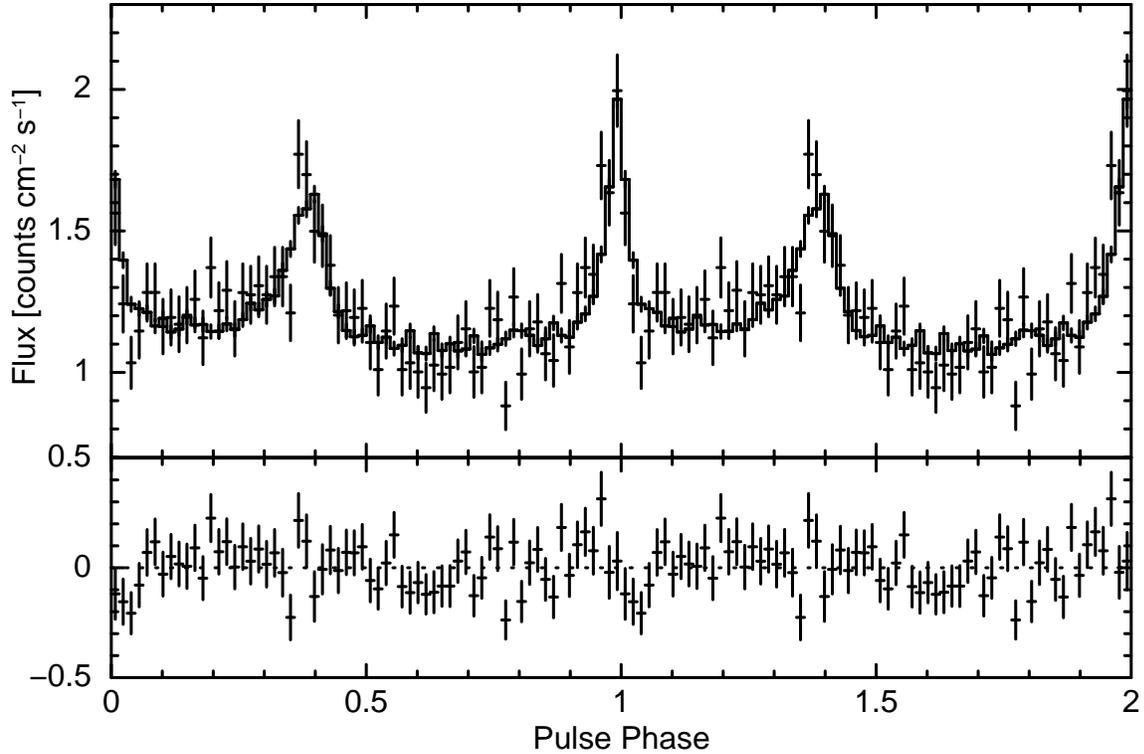}
  \end{center}
  \caption{Top panel: Pulse profile of the Crab pulsar in the 4$-$10 keV
    during the GeV gamma-ray flare for Period A+B (MJD 55457.5$-$55462.5) obtained by the GSC.
    The vertical axis is the flux shown in a unit of counts cm$^{-2}$ s$^{-1}$.
    The horizontal axis is the pulse phase.
    The solid histogram is the template pulse profile
    with the best-fit parameters (see text).
    The bottom panel is the residual of the data from the best-fit model shown
    in the same unit with the top panel.
    In both panels, the vertical error bars correspond to 1-$\sigma$ statistical errors.
    The same profiles are shown in two cycles.
}\label{fig: pulse_flare}
\end{figure}

We repeated the same analysis from 2009 November 1 to 2010 November 29
every 2.5 days to make the light curve of the pulsed flux in the 4$-$10 keV band
[Figure \ref{fig: lc} (bottom panel)] and
fitted it by a constant.
The reduced chi-squared of the fit is 1.27 for 104 DOF, meaning that
there was no evidence of variability.
The flux obtained is shown in table \ref{model_values}.
The flux variations observed during Periods A, B, and A+B are statistically
consistent with the best fit function.
The statistical 90\% confidence level upper limits on the variation
of the pulsed fluxes for these periods
are 9.0, 37.3, and 18.8\%, respectively (table \ref{model_values}).
The 0.5-day binned light curve of the pulsed flux around the flare period
is shown in figure \ref{fig: lc_flare} (bottom panel).
The variation from the best-fit constant function ``Pulsed: const'' of table \ref{model_values}
is not statistically significant with a reduced chi-squared of 1.44 for 12 DOF.

\begin{table}
  \caption{Pulse phase averaged and pulsed fluxes of the Crab nebula in $4-10$ keV.
}\label{flux_values}
  \begin{center}
    \begin{tabular}{lccc}
      \hline \hline
     Flux$^*$ & Period A$^\dagger$  & Period B$^\ddagger$ & Period A+B$^{\S}$\\ 
    \hline
    Averaged  & $1.21\pm0.02$ & $1.23\pm0.02$ & $1.22\pm0.01$ \\
    Pulsed    & $0.13\pm0.02$ & $0.18\pm0.02$ & $0.16\pm0.01$ \\ \hline
    \multicolumn{4}{l}{$^*$photons cm$^{-2}$ s$^{-1}$ with 1-$\sigma$ statistical error.}\\
    \multicolumn{4}{l}{$^\dagger$MJD 55457.5$-$55460.0. $^\ddagger$MJD 55460.0$-$55462.5.}\\
    \multicolumn{4}{l}{$^{\S}$MJD 55457.5$-$55462.5}\\
    \end{tabular}
  \end{center}
\end{table}

\begin{longtable}{*{5}{l}}
\caption{
Model parameters and upper limits on the variation of the flux
of the Crab nebula in $4-10$ keV
}\label{model_values}
\hline\hline
Model        &   & Average: linear$^*$ (free) & Average: linear$^*$ (fixed)
             & Pulsed: const \\
\endfirsthead
\hline
\multicolumn{5}{l}{$^*$The linear function is
 $F(t) = F(t_{\rm mid}) \left[1 + S \times (t - t_{\rm mid})\right]$, where $F(t)$, $t$, $t_{\rm mid}$
 and $S$ are the flux}\\
\multicolumn{5}{l}{(photons cm$^{-2}$ s$^{-1}$), time (day), mid-time (MJD 55312.5)
and the slope (day$^{-1}$), respectively.}\\
\multicolumn{5}{l}{$^{\dagger}$At the mid-time with 1-$\sigma$ statistical error.}\\
\multicolumn{5}{l}{$^{\ddagger}$Statistical 90\% confidence level.
$^{\S}$Same as table \ref{flux_values}.}\\
\endlastfoot
\hline
Parameters   & Flux$^{\dagger}$  & $1.241(2)$ & $1.242(1)$
             & $0.145(1)$ \\ 
             & Slope$^{\dagger}$  & $(-8.5\pm1.0) \times 10^{-5}$ & $-7.2 \times 10^{-5}$(fix) 
             & --- \\ \hline
Upper limits$^{\ddagger}$ & Period A$^{\S}$  & 0.8\% & 0.6\% 
             & 9.0\% \\
             & Period B$^{\S}$ & 2.4\% & 2.1\%
             & 37.3\% \\
             & Period A+B$^{\S}$ & 1.0\% & 0.8\%
             & 18.8\% \\
\end{longtable}

\section{Conclusion}\label{sec: conclusion}

We report on the MAXI GSC observation of the Crab nebula during the GeV gamma-ray flare.
We successfully detected the pulsation of the Crab pulsar during the simultaneous period,
and conclude that there is no evidence for changes in the pulse profile,
pulsed flux and pulse phase averaged flux
during the gamma-ray flare.
We obtained an upper limit on the variation of the pulse phase averaged flux
of $1$\% at a 90\% confidence limit of the statistical uncertainty
from the best-fit linear function during the 5 day interval of
the gamma-ray flare in the 4$-$10 keV band.
During the same period in the same energy band,
we also obtained an upper limit on the variation of the pulsed flux
of $19$\% at a 90\% confidence limit of statistical uncertainty
from the mean level.

The MAXI GSC simultaneous observation with the gamma-ray flare
is uniquely important to constrain the origin of the flare,
in contrast to the follow-up observations 
performed after the cease of the gamma-ray flare.
The lack of changes on the pulsed component in the X-ray (this work),
as well as those in radio \citep{Espinoza+2010},
hard X-ray (Super-AGILE observation at the flare on 2007 shown in \citet{Tavani+2011})
and gamma-ray bands \citep{Tavani+2011, Abdo+2011},
supports the nebular origin for the gamma-ray flare as proposed
in several papers \citep{Tavani+2011, Abdo+2011, Bednarek Idec 2010,
Yuan+2010, Komissarov Lyutikov 2010}.
In spite of the large flux increase of factor five in the GeV energy region \citep{Abdo+2011},
we constrain a limit on the variation of the nebula flux in the X-ray band.
This provides valuable information to construct theoretical models for the gamma-ray flare of the Crab nebula.

\bigskip

We are grateful to the members of the MAXI operation team.
We acknowledge the use of the Crab ephemeris provided at the web site of
the Jodrell Bank Centre for Astrophysics \citep{Lyne+1993}.
This research was partially supported by the Ministry of Education, Culture, Sports, Science and Technology (MEXT),
Grant-in-Aid No.19047001, 20041008, 20540230, 20244015 , 20540237, 21340043, 21740140, 22740120,
and Global-COE from MEXT ``The Next Generation of Physics, Spun from Universality and Emergence''
and ``Nanoscience and Quantum Physics.''

\end{document}